\documentstyle[sprocl]{article}

\input{psfig}
\def\be{\begin{equation}}
\def\ee{\end{equation}}
\def\bea{\begin{eqnarray}}
\def\eea{\end{eqnarray}}

\begin{document}

\title{ON THE SPECTRUM OF THE QCD DIRAC OPERATOR}

\author{ J.J.M. VERBAARSCHOT}

\address{Department of Physics, \\
SUNY at Stony Brook,\\ Stony Brook, NY 11794}


\maketitle\abstracts{
In this lecture we argue that the fluctuations of Dirac 
eigenvalues on the finest scale, i.e. on the scale of the average
level spacing do not depend on the underlying dynamics and can be 
obtained from a chiral random matrix theory  with the same low
energy effective theory. We present three pieces of evidence
supporting that such microscopic correlations  of 
lattice QCD Dirac spectra are given by chiral random 
matrix theory. First, we find that the spectral correlations of eigenvalues
in the bulk of the spectrum obey the Dyson-Mehta-Wigner statistics. Second, we
show that the valence quark mass dependence for sufficiently small quark
masses, as calculated by the Columbia group,   
can be obtained from the microscopic spectral density of chiral random matrix
theory. Third, in the framework of chiral 
random matrix models, we present results showing
that the microscopic spectral density is strongly universal, 
i.e. is insensitive to the details of the probability distribution}.

\section{Introduction}

Although not a physical observable itself, the spectrum of the 
Euclidean QCD Dirac
operator is an essential ingredient for the calculation of hadronic 
correlation functions. It also serves as the order parameter for the 
chiral phase transition. This becomes clear by expressing the quark propagator
in terms of the eigenfunctions and the eigenvalues of
the Dirac operator
\be
S(x,y) = -\sum_k \frac { \phi_k(x) \phi_k^*(y)}{\lambda_k + im},
\ee
where 
\be
i\gamma D \phi_k = \lambda_k \phi_k,
\ee
and $i\gamma D$ is the Euclidean Dirac operator. 
Because $\{\gamma_5, i\gamma D\} = 0$, all nonzero 
eigenvalues occur in pairs $\pm\lambda_k$.  	
The
eigenvalues near zero are of great importance for the propagation of
light quarks. This is the reason that instanton field configurations,
with one zero eigenvalue per instanton, determine the 
main characteristics of hadronic correlation functions \cite{SV}. 

With regards to chiral symmetry breaking, the Dirac operator is directly
related to the chiral condensate
\bea
\langle \bar q q\rangle &=& -\lim \frac 1V \int d^4 x S(x,x),\nonumber \\
&=& \lim \frac 1V \int \frac {2m i\rho(\lambda) d\lambda}{m^2 
+\lambda^2},\nonumber \\ 
&=& \frac {\pi i\rho(0)}{V},
\eea
which is the celebrated Banks-Casher formula \cite{Banks-Casher}.
Here, the spectral density is defined as
\be
\rho(\lambda) = \sum_k \langle \delta(\lambda-\lambda_k) \rangle,
\ee
and the average is over gauge field configurations weighted according to
the Euclidean action. The space-time volume is denoted by $V$. The limit
is defined such that the thermodynamic limit is taken before the chiral limit.

In this lecture, I wish to argue that some properties of the Dirac spectrum
are completely determined by the global symmetries of the QCD partition 
function and are not sensitive to the dynamics of the theory. In general
such universal properties are given by fluctuations of the eigenvalues on the
finest scale, i.e. on the scale of individual level spacings. 
Because the spectrum is symmetric about zero, we have to distinguish spectral
correlations near zero virtuality from correlations in the bulk of the 
spectrum. Although the overall spectral density is certainly not universal,
we wish the convince the reader that the spectral density near zero, on the
scale of the average level spacing, is universal \cite{SVR,V,VZ}. 
It is called the
microscopic spectral density. In the bulk of the spectrum, the 
spectral correlations will be shown to obey universal level statistics
\cite{HV}.
From the study of quantum
chaos, we know that the latter correlations are strongly universal
\cite{univers,bz}. In the
framework of a random matrix model we will show that the microscopic spectral
density has strong universality properties as well \cite{zee,sener}.
Finally, we make contact with lattice QCD calculations. In this context
the  Dirac spectrum was investigated via the valence quark mass dependence
of chiral condensate \cite{Christ}, and, for a relatively small number
of configurations,  by  a full diagonalization of
the Dirac operator \cite{Teper,Kalkreuter}. 
Both the valence quark mass dependence \cite{vplb} 
and the eigenvalue correlations
in the bulk of the spectrum \cite{HV} are shown to be in complete agreement
with random matrix theory (RMT) within its range of applicability.

\section{Universality}

Leutwyler and Smilga \cite{LS} have argued that in the range

\be
\frac 1\Lambda \ll L\ll \frac 1{\sqrt {m\Lambda}},
\label{range}
\ee
where $L$ is the linear size of the Euclidean box, $m$ the quark mass and
$\Lambda$ a typical hadronic mass scale,
the mass dependence 
of the QCD partition function is given by the effective partition function
\be
Z_{\rm eff}(M, V) = \int_{U\in G/H} dU \exp({\rm Re} \,V\Sigma 
\,{\rm tr}\, M U e^{i\theta/N_f}).
\label{zeff}
\ee
Here, $M$ is the mass matrix, $\theta$ is the vacuum angle
and $\Sigma= |\langle \bar q q \rangle |$. 
The integration  is over the Goldstone manifold $G/H$ which is
determined by the scheme of chiral symmetry breaking. In QCD with three
colors and $N_f$ fundamental fermions, $G/H = SU(N_f)$.

However, QCD is not the only theory that can be reduced to the 
effective partition function (\ref{zeff}). In particular, chiral
random matrix theories can be reduced to this partition 
function \cite{SVR}. This leads us to the notion of 
universality:
a property will be called universal if it is the same for theories 
with the same low energy effective theory. This allows us to calculate such 
quantities for the simplest possible theory, for example, for a chiral random
matrix theory. 
In particular, properties that are determined by the effective partition
function are trivially universal. As an example, I mention the 
Leutwyler-Smilga sum rules. Let us for the moment consider the sector
of zero topological charge. Then, by expanding both sides of
the equality (valid in the range (\ref{range}))
\be
\frac{Z^{\rm QCD}_{\nu=0}(m,V)}{Z^{\rm QCD}_{\nu=0}(m=0,V)} =
\frac{Z^{\rm eff}_{\nu=0}(m,V)}{Z^{\rm eff}_{\nu=0}(m=0,V)}
\ee
in powers of $m$, and performing the  group integrals we obtain an 
infinite family of sum rules. The simplest sum rule is given by
\cite{SmV}
\be
\frac 1{V^2} \sum_{\lambda_k > 0} \left \langle \frac 1{\lambda_k^2} 
\right\rangle_{\nu=0}
= \frac{\Sigma^2}{4 N_f}.
\ee
The generalization of this formula to a sector with topological charge $\nu$
and different coset spaces
($G/H = SU(2N_f)/Sp(2N_f)$ for $SU(2)$ color with fundamental fermions,
$G/H = SU(N_f)/SO(N_f)$ for adjoint fermions with gauge group $SU(N_c), N_c \ge 
2$) is given by
\be
\frac 1{V^2} \sum_{\lambda_k > 0} \left\langle 
\frac 1{\lambda_k^2} \right\rangle_{\nu}
= \frac {\Sigma^2}{4 \left (|\nu| + ({\rm dim (coset)} + 1)/N_f\right )}.
\label{sumrule}
\ee

The sum  in (\ref{sumrule}) can be replaced by an integral over the spectral
density
\be
\frac 1{V^2} \sum_{\lambda_k > 0} \left\langle 
\frac 1{\lambda_k^2} \right\rangle_{\nu}
= \Sigma^2 \int_0^\infty \frac {du}{u^2} 
\frac 1{V\Sigma} \rho(\frac u{V\Sigma}),
\ee
where we have introduced the microscopic variable $u = \lambda V \Sigma$.
The combination
\be
\rho_S(u) \equiv \frac 1{V\Sigma} \rho(\frac u{V\Sigma}),
\ee
will be called the microscopic spectral density. It is constrained by
the effective partition function but not determined by it. Our conjecture
is that, in the thermodynamic limit, it is universal function 
that can be calculated with the help
of chiral random matrix theory.

A more intuitive interpretation of $\rho_S(u)$
is obtained if we start from the Banks-Casher
relation and the fact that chiral symmetry is broken. The eigenvalues near
zero virtuality are roughly equally spaced, with spacing given by
\be
\Delta \lambda = \frac 1{\rho(0)} = \frac {\pi}{\Sigma V}.
\ee
and smallest nonzero eigenvalue of the Dirac operator is equal to 
\be
\lambda_{\rm min} \sim \frac 1{V\Sigma}.
\ee
The microscopic variable $u$ parameterizes the spectral density on the
scale of the average level spacing, and the microscopic spectral density
is obtained by magnifying the region around $\lambda=0$ by a factor $V$. 

An observable directly related to the microscopic spectral density is
the valence quark mass dependence of the chiral condensate \cite{Christ}
which can be expressed as
\be
\frac {\Sigma(m)}\Sigma =\int_0^\infty \frac {2mV\Sigma}{u^2 +m^2V^2 \Sigma^2}
\rho_S(u) du.
\label{valence}
\ee
It can be obtained from random matrix theory in the range (\ref{range}).
This range (\ref{range}) can also be written as
\be
m \ll \sqrt{\lambda_{\rm min} \Lambda},\qquad
\lambda_{\rm min} \ll \Lambda,
\ee
which might be called the mesoscopic window of QCD.

To summarize this section,  we conclude that the eigenvalues of the QCD Dirac
operator are strongly correlated for the following reasons:

\begin{itemize}
\item The Leutwyler-Smilga sum-rules.
\item The eigenvalue spacing near zero $\Delta \lambda \sim 1/V$ instead of
$\Delta \lambda \sim 1/V^{1/4}$ for a noninteracting system.
\item If the eigenvalues where uncorrelated, then, in the chiral limit
      $\rho(\lambda) \sim \lambda^{2N_f}$, and chiral symmetry would not 
be broken.
\end{itemize}

\section{Universal level correlations of quantum spectra}

Quantum spectra of complex systems have been investigated in great detail
both  experimentally
and theoretically. Already, in the fifties it was observed that
the spectral correlations of nuclear resonances \cite{Wigner} do not depend
on the details of the system and can be obtained from random matrix theory.
More recently, it has been shown that the essential ingredient to obtain
random matrix spectra is that the corresponding classical system is
chaotic \cite{univers}. Even systems with only two degrees of freedom
show spectral correlations that are in perfect agreement with
the random matrix prediction. It was already noted early on that the average
spectral density is not given by random matrix theory. Typically, for
a physical system, it is a monotonously increasing function whereas for 
random matrix theory it is a semicircle. To deal with this problem,
a first important hypothesis was introduced, namely, that we have
a separation of scales with regards to
the average spectral density and the microscopic
spectral fluctuations. The second hypothesis is that the microscopic 
fluctuations are given by the random matrix ensemble with same $symmetries$
as the underlying theory. In particular, anti-unitary symmetries play an 
important role in identifying the correct RMT.

In practice, the average spectral density is folded out by transforming
the original spectrum $\{\lambda_k\}$ to a new spectrum with average
spectral density equal to one. Below we will use the following
statistics to measure the spectral correlations of the unfolded level
sequence: the nearest
neighbor spacing distribution, $P(S)$, the number statistics and the
$\Delta_3$-statistic.
The number statistics are obtained by counting the number of eigenvalues 
$n_k$ in consecutive
intervals of length $n$. The number variance is defined as
$\Sigma_2(n) = var\{n_k\}$, and
the $\Delta_3$-statistic
is obtained from a convolution of $\Sigma_2(n)$ with a 'smoothening'-kernel.

The above statistics have been obtained analytically (see \cite{Wigner})
for the invariant random
matrix ensembles defined by a matrix of independently distributed
gaussian matrix elements, consistent with the
hermiticity of the matrix. Depending on the anti-unitary symmetry, the
matrix elements are real (Gaussian Orthogonal Ensemble (GOE), $\beta = 1$), 
complex (Gaussian Unitary Ensemble (GUE), $\beta = 2$) 
or quaternion real (Gaussian Symplectic Ensemble (GSE), $\beta = 4$). 
The most notable property is the stiffness of
RMT spectra: the number variance behaves asymptotically as
$\Sigma_2(n) \sim \frac 2{\pi^2\beta} \log (n)$ instead of 
$\Sigma_2(n) \sim n$ for independently distributed eigenvalues.

\section{Chiral random matrix theory}

Chiral random matrix theories are theories with the global symmetries of
the QCD Dirac operator,  but otherwise independently
Gaussian distributed random matrix elements. 
The chiral random matrix model that obeys these conditions is defined by
\cite{SVR,V,VZ}
\be
Z_\nu^\beta = \int DW \prod_{f= 1}^{N_f} \det({\cal D} + m_f)
\exp(-\frac{N\Sigma^2 \beta}4 {\rm Tr}W^\dagger W),
\label{model}
\ee
where
\be
{\cal D} = \left (\begin{array}{cc} 0 & iW+i(\Omega_T -arg(P))+\mu\\ 
iW^\dagger +i( \Omega_T -arg(P))+\mu)
 & 0 \end{array} \right ),
\label{zrandom}
\ee
and $W$ is a rectangular  $n\times m$ matrix with $\nu = n-m$ and
$N= n+m$. An early version of this model can be found in \cite{early}.
The temperature dependence introduced in \cite{JV,Stephanov1}
is included in the form of Matsubara
frequencies in the diagonal matrix $\Omega_T$. 
The chemical potential, $\mu$, introduced in \cite{Stephanov2}
results in a non-Hermitean Dirac operator with eigenvalues that
are scattered in the complex plane. The term
$\arg(P)$, the argument of the Polyakov loop was introduced by Stephanov
\cite{Stephanov1} in order to explain the dependence of the critical 
temperature on the expectation value of the $Z_N$-phase \cite{Christ}. 
In this paper we only discuss models with $\mu = 0$ and
temperature dependence with only the lowest Matsubara frequencies included.
This partition function
reproduces the following symmetries of QCD:

\begin{itemize}

\item The $U_A(1)$ symmetry. All nonzero eigenvalues of the random matrix
Dirac operator occur in pairs $\pm \lambda$.

\item The topological structure of the QCD partition function. The matrix
${\cal D}$ has exactly $|\nu|\equiv |n-m|$ zero eigenvalues. This identifies
$\nu$ as the topological sector of the model. 

\item The flavor symmetry is the same as in QCD. For $\beta = 2$ it is
$SU(N_f) \times SU(N_f)$. For $\beta = 1$ it is $SU(2N_f)$ and for
$\beta = 4$ it is $SU(N_f)$.

\item The chiral symmetry is broken spontaneously for two or more flavors
according to the pattern $SU(N_f) \times SU(N_f)/SU(N_f)$,
$SU(2N_f)/Sp(N_f)$ and $SU(N_f)/O(N_f)$ for $\beta = 2$, 1 and 4,
respectively, the same as in QCD \cite{Shifman-three}.
The chiral condensate satisfies the Banks-Casher relation
\be
\Sigma = \partial_{m_f} \log Z = 
\lim_{N\rightarrow \infty} \frac {\pi \rho(0)}N.
\ee
Therefore, the parameter $\Sigma$ in the random matrix model is identified
as the chiral condensate and $N$ as the (dimensionless) volume of space
time.

\item The anti-unitary symmetries. For three and more colors with
fundamental fermions the Dirac operator has no anti-unitary symmetries,
and, in general, the matrix elements of the Dirac operator are
complex. The matrix elements of $W$ of the corresponding random matrix
ensemble are chosen arbitrary complex as well ($\beta =2$).
For $N_c =2$ and fundamental fermions the Dirac operator satisfies
\be
[C\tau_2 K, i\gamma D] = 0,
\ee
where $C$ is the charge conjugation matrix and $K$ is the complex conjugation
operator.
Because, $(C\tau_2 K)^2 =1$, the matrix elements of the Dirac operator
can always be chosen real, and the corresponding random matrix ensemble
is defined with real matrix elements ($\beta = 1$).
For two or more colors with fermions
in
the adjoint representation $i\gamma D$ has the symmetry
\be
[CK, i\gamma D] =0,
\ee 
but now $(CK)^2 = -1$, which allows us to rearrange the matrix elements of
the Dirac operator into real quaternions. The matrix elements of $W$ of the
corresponding random matrix ensemble are chosen quaternion real as well
$(\beta =4)$.

\end{itemize}
                  
\section{Universal spectral correlations}

As already stated before, we have to distinguish spectral correlations near 
zero and spectral correlations in the bulk of the spectrum. The latter have 
the advantage that, under the assumption of spectral ergodicity, they can
be calculated by a spectral average rather than an ensemble average. This
requires no more than a few statistically independent gauge field 
configurations, whereas the calculation of the microscopic spectral density
requires a prohibitively large number of independent gauge field 
configurations. 

We will present three pieces of evidence that spectra of the lattice QCD
Dirac operator show universal correlations on the finest scale. First,
we will show that spectral correlations in the bulk of the spectrum are given 
by the invariant random matrix ensembles. Second, we will show that the
valence quark mass dependence in the mesoscopic range of QCD can be obtained
from the microscopic spectral density. Third, we will present random matrix
results showing that that microscopic spectral density is insensitive
to the probability distribution of the matrix elements and to the
temperature as introduced in (\ref{model}).

\subsection{Correlations in the bulk of lattice spectra}

Recently, complete Dirac spectra for reasonably large lattices were
obtained by Kalkreuter \cite{Kalkreuter}
both for Kogut-Susskind fermions (with four flavors)
and Wilson fermions (with two flavors) with $SU(2)$-color.
It is important that $all$ eigenvalues
were obtained. Because the sum of the squares of the eigenvalues satisfy
rigorous sum-rules, this allows a very stringent test on the accuracy
of the calculation.

In the case of the $SU(2)$ color group, the anti-unitary symmetries of
Kogut-Susskind fermions are different from the continuum theory. We have have
\be
[\tau_2 K,  D^{KS}] = 0 \quad {\rm with} \quad (\tau_2 K)^2 = -1,
\ee
($K$ is the complex conjugation operator) whereas for Wilson fermions in the
form considered by Kalkreuter we have
\be
[\gamma_5 CK\tau_2, \gamma_5 \gamma D^{\rm Wilson}] = 0 \quad {\rm with}
(\gamma_5 CK\tau_2)^2 = 1.
\ee
According to random matrix theory \cite{Wigner}, 
if the square of the anti-unitary operator
is $-1$ the spectral correlations are given by the GSE, whereas, if the
square is $1$, the correlations are given by the GOE. 
\begin{figure}
\psfig{figure=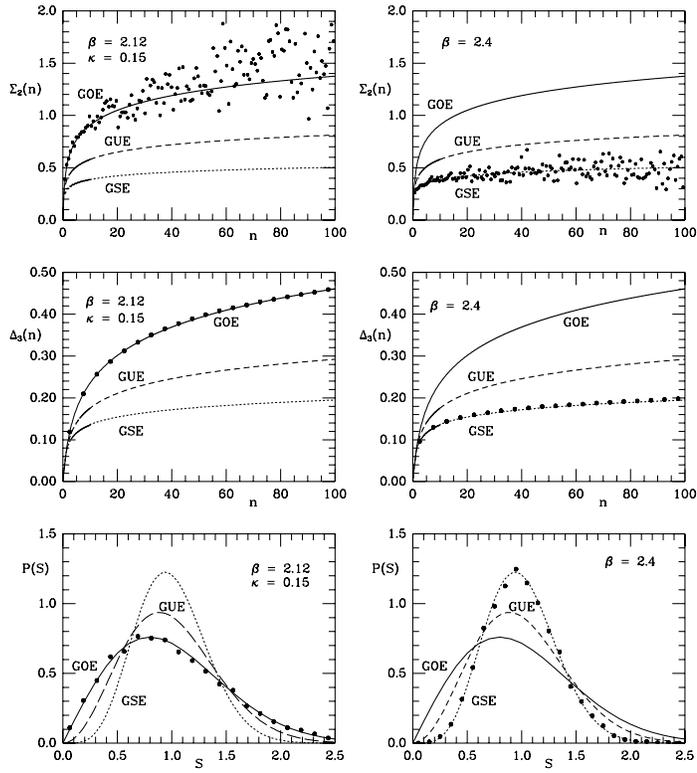,height=4in}
\caption{Spectral correlations of Dirac eigenvalues for Wilson fermions
(left) and Kogut-Susskind fermions (right). The number variance, $\Sigma_2(n)$,
the $\Delta_3$-statistic and the nearest neighbor spacing distribution
are shown in the upper, middle and lower figure, respectively.}
\end{figure}

We have analyzed the spectra of a score of lattice QCD configurations. In Fig.
1 we show the results for the correlations of the Dirac eigenvalues for
Kogut Susskind fermions (right) on a $12^4$ lattice and Wilson fermions (left)
on a $8^3 \times 12$ lattice for parameters as indicated in labels of 
the figure.
We show the number variance, $\Sigma_2(n)$, 
the $\Delta_3$ statistic and the nearest neighbor spacing distribution
$P(S)$. The points represent the lattice-data and the random matrix results
are given by the curves. The corresponding RMT is given in the label of the 
figure. 
We find perfect agreement with the above
predictions. In addition, we have also confirmed 
that the correlations are stationary
over the spectrum. Even if we consider only the first 100 or 200 eigenvalues,
no deviations from RMT correlations are found.
 
\subsection{The valence quark mass dependence of the chiral condensate.}

Using standard methods of random matrix theory it is possible to obtain
analytically the microscopic spectral density. The result for
$N_f$ flavors in the sector of topological charge $\nu$ is \cite{V}
\be
\rho_S(u) = \frac u2 \left ( J^2_{N_f+\nu}(u) - 
J_{N_f+\nu+1}(u)J_{N_f+\nu-1}(u)\right).
\ee
Using (\ref{valence}) we obtain the valence quark mass dependence by an
elementary integration \cite{vplb}
\be
\frac {\Sigma(x)}{\Sigma} = x(I_{N_f+\nu}(x)K_{N_f+\nu}(x)
+I_{N_f+\nu+1}(x)K_{N_f+\nu-1}(x)).
\label{val}
\ee
Asymptotically, for large $x\equiv mV\Sigma$, $\Sigma(x) \sim 
\Sigma(1-(N_f+\nu)/x)$
and for $x \rightarrow 0$, $\Sigma(x) \sim x\Sigma/2(N_f+\nu)$ for
$N_f+\nu\ne 0$ and $\Sigma(x) \sim -x \Sigma \log(x)$
for $N_f+\nu = 0$.
\begin{figure}
\psfig{figure=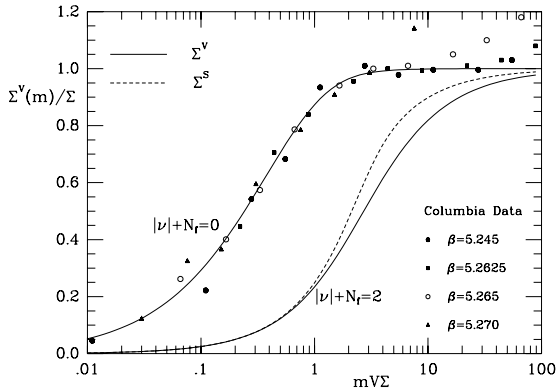,height=2.0in,angle=270}
\caption{The valence quark mass dependence of the chiral condensate
$\Sigma^V(m)$ plotted as $\Sigma^V(m)/\Sigma$ versus $m N\Sigma$. The dots and
squares represent lattice results by the Columbia group [11] for
values of $\beta$ as indicated in the label of the figure.}
\end{figure}
The valence quark mass was calculated by the Columbia group \cite{Christ}
for unquenched Kogut-Susskind fermions on
a $16^3\times 4$ lattice with $m_{\rm sea} a = 0.01$, two flavors and three 
colors. 
In Fig. 2 we plot their results in terms of the dimensionless
quantities suggested by the valence quark mass dependence (\ref{val}), namely
as $\Sigma(x)/\Sigma$ versus $x$. We observe that the lattice
data fall on a universal curve given by (\ref{val}) for $N_f + \nu = 0$. 
The reason that $N_f= 0$ is that $\lambda_{\rm min} a \sim 1.0e-4 \ll m_{\rm 
sea} a$, so that the calculation is quenched on the scale of the smallest
eigenvalues. It is also not surprising that $\nu =0$ because the fermionic
zero modes are completely mixed with the rest of the states. For larger
$x$ the lattice data deviate from the universal curve. This was to be
expected because in terms of $x$ the domain of validity is given by
\be
x \ll \sqrt{\frac \Lambda{\lambda_{\rm min}}}.
\ee

\subsection{Universality in random matrix theory}

The microscopic spectral density was calculated for the simplest possible
random matrix ensemble. Recently, the microscopic spectral density
for  $N_f = \nu =0$ was calculated for ensembles with a nongaussian
probability distribution \cite{zee}. The authors found that
the addition of a polynomial in $tr (W^\dagger W$) to the exponent of
the probability distribution had no effect on the microscopic spectral density.
We studied the random matrix model (\ref{zrandom}) 
with only the lowest Matsubara 
frequencies included \cite{sener}. 
It can be shown that it is sufficient to include only the 
positive frequency. This model has a critical temperature, $T_c$,  above which
$\rho(0)= 0$. In spite of the fact that the average spectral density
changes drastically between $T=0$ and $T=T_c$  (it is
given by the solution of a cubic equation \cite{JV,Stephanov1})
we find that the microscopic spectral density remains unaffected.

Numerically, we studied  a variety other ensembles. In particular, I 
mention the chiral Cauchy ensemble, with a with a diverging second moment
of the matrix elements. Nevertheless, we find the same universal
microscopic spectral density. 
Finally, we wish to mention that according to general universality arguments
in random matrix theory, the correlations in the bulk of the spectrum
of the chiral random matrix ensembles
are given by the invariant random matrix ensembles \cite{bz}.
Therefore, we find that the microscopic spectral density and the 
correlations in the bulk of the spectrum are mutually inclusive
and both are strongly universal.
\section {Conclusions} 
We have argued that the spectral fluctuations on the finest scale are 
determined by the global symmetries of the QCD partition function, and can
be obtained from chiral random matrix theories. In particular, the microscopic
spectral density determines the valence quark mass dependence of the
condensate in the mesoscopic range of QCD. In this range we find perfect
agreement with recent lattice QCD data. We have also observed that correlations
in the bulk of the spectrum are given by the invariant random matrix ensembles.
This implies that the spectral fluctuations are strongly suppressed
$\Sigma_2(n) \sim \log(n) /\pi^2$, instead of $n$ for uncorrelated eigenvalues.
To some extent, QCD is self-quenching.

\section*{Acknowledgments}

 The reported work was partially supported by the US DOE grant
DE-FG-88ER40388.

\end{document}